\documentstyle[twocolumn,prl,aps,epsfig,eufrak]{revtex}

\begin{document}

\def\4he{$^4$He}
\def\pc{\protect\cite}
\def\hpsi{\hat\psi}
\def\tpsi{\tilde\psi}
\def\br{{\bf r}}
\def\bk{{\bf k}}
\def\bu{{\bf u}}
\def\bw{{\bf w}}
\def\brt{\br,t}
\def\bbrt{(\brt)}
\def\cphio{\Phi_0}
\def\beq{\begin{equation}}
\def\eeq{\end{equation}}
\def\bea{\begin{eqnarray}}
\def\eea{\end{eqnarray}}
\def\bna{\bbox{\nabla}}
\def\bp{{\bf p}}
\def\bv{{\bf v}}
\def\tn{\tilde n}
\def\tp{\tilde p}
\def\be{\bbox{\eta}}
\def\imag{{\rm{Im}}}
\newcommand{\eq}[1]{\mbox{(\ref{eq#1})}}

\wideabs{ 
\title{Damping of condensate collective modes due
to equilibration with the non-condensate}

\author{J. E. Williams and A. Griffin}

\address{Department of Physics, University of Toronto, Toronto,
Ontario M5S 1A7, Canada}

\date{\today}

\maketitle

\begin{abstract}
We consider the damping of condensate collective modes in the
collisionless regime at finite temperatures arising from lack of
equilibrium between the condensate and the non-condensate atoms, an
effect that is ignored in the usual discussion of the collisionless
region. As a first approximation, we ignore the dynamics of the
thermal cloud. Our calculations should be applicable to collective
modes of the condensate which are oscillating out-of-phase with the
thermal cloud. We obtain a generalized Stringari equation of motion
for the condensate at finite temperatures, which includes a damping
term associated with the fact that the condensate is not in diffusive
equilibrium with the static thermal cloud. This inter-component
collisional damping of the condensate modes is comparable in magnitude
to the Landau damping considered in the recent literature.

\vspace{.2cm} PACS numbers(s): 03.75.Fi.~~05.30.Jp~~67.40.Db
\end{abstract}
}

\section{Introduction}

The collective oscillations of a condensate at zero temperature $T=0$
are well described by the solutions of the linearized Gross-Pitaevskii
(GP) time-dependent equation of motion for the condensate wavefunction
$\Phi(\br,t)$. At finite temperatures, the condensate dynamics is
modified by interactions with the non-condensate atoms in the thermal
cloud, which has the effect of renormalizing and damping the
condensate oscillations.  Recently the coupled dynamics of the
condensate and the thermal cloud has been the subject of several
theoretical
studies~\cite{Stoof99,Walser99,Zaremba99,Gardiner2000,Milena2000}. Such
calculations lead to a generalized GP equation for $\Phi(\br,t)$ and
some appropriate Boltzmann-like kinetic equation describing the
dynamics of the non-condensate atoms.  In the present work, we make
use of the recent formulation of Zaremba, Nikuni, and Griffin
(ZNG)~\cite{Zaremba99} to discuss a new kind of damping of
condensate oscillations that arises from collisions between the
condensate and non-condensate components. In contrast to
Ref.~\cite{Zaremba99}, which discussed the collision-dominated
hydrodynamic regime, here we discuss the collisionless regime. Within
the well-known Thomas-Fermi (TF) approximation, we derive a
generalized Stringari wave equation describing the condensate normal
modes~\cite{Stringari96} that is valid at finite $T$ and includes
damping due to the fact that the condensate is not in equilibrium with
the thermal cloud. This new source of damping is in addition to the usual
Landau and Beliaev damping considered in the collisionless region at
finite
$T$~\cite{Szepfalushy74,Shi97,Pitaevskii97,Fedichev98a,Giorgini98,Guilleumas2000,Morgan2000,Reidl2000}.

Our theory can be used to generalize any discussion based on the usual
GP equation at $T=0$. This simplicity is due to our neglect of any
dynamics of the thermal cloud. Available studies of collective
modes~\cite{Zaremba99} at finite $T$ suggest that, for any given mode
symmetry, one mode mainly involves motion of the condensate (with a
small out-of-phase motion of the non-condensate). This mode is a
natural extension of the $T=0$ oscillation of a pure condensate and
should be described by our theory. The other mode, of the same
symmetry, mainly involves the motion of the thermal cloud (with a
small in-phase motion of the condensate) and can be viewed as the
natural extension of the oscillations above the critical temperature
$T_{\rm{BEC}}$~\cite{Kavoulakis98,Guery-Odelin1999}.  Our present
calculations do not apply to such ``normal fluid'' oscillations, 
which include the Kohn mode at the trap frequency.

\section{Derivation of Model}

Our starting point is the finite $T$ generalized GP equation derived
by ZNG (see also Refs.~\cite{Stoof99} and \cite{Walser99})
\beq i\hbar {\partial \Phi \over \partial t} = \Big [ - {\hbar^2 \over
2 m} \nabla^2 + U_{\rm{ext}} + g n_c + 2g \tilde{n} - i \hbar R \Big ]
\Phi ,
\label{eq1}
\eeq where the interaction parameter $g = 4 \pi \hbar^2 a/m$, $a$ is
the s-wave scattering length, $n_c(\br,t) = |\Phi(\br,t)|^2$, and
$\tilde{n}(\br,t)$ is the non-condensate local density. The damping
term in \eq{1} is given by $R(\br,t) \equiv \Gamma_{12}(\br,t) /2
n_c(\br,t)$, with \beq \Gamma_{12}(\br,t) = \int {d\bp \over (2 \pi
\hbar)^3} C_{12}[f(\bp,\br,t),\Phi(\br,t)].
\label{eq2}
\eeq This involves the collision integral $C_{12}[f,\Phi]$ describing
collisions of condensate atoms with the thermal atoms, which also enters
the approximate semi-classical kinetic equation for the
single-particle distribution function (valid for $k_{\rm{B}} T \gg g
n_{c0}$ and $k_{\rm{B}} T \gg \hbar \omega_0$) \beq {\partial f \over
\partial t} + {\bp \over m} \cdot \bbox{\nabla} f- \bbox{\nabla} U
\cdot \bbox{\nabla}_{\bp} f = C_{12}[f,\Phi] + C_{22}[f,\Phi]\,.
\label{eq3}
\eeq Here the collision integral $C_{22}[f]$ describes binary
collisions between non-condensate atoms. It does not change the number
of condensate atoms and hence does {\emph{not}} appear explicitly in
\eq{1}. These coupled equations~(\ref{eq1})-(\ref{eq3}),
{\mbox{along}} with equations (23a) and (23b) of Ref.~\cite{Zaremba99}
defining the collision integrals $C_{12}$ and $C_{22}$, were derived
in the semi-classical approximation. However, they are expected to
contain all the essential physics in trapped Bose-condensed gases at
finite $T$, in both the collisionless and hydrodynamic domains. They
assume that the atoms in the thermal cloud are well-described by the
single-particle Hartree-Fock spectrum $\tilde \varepsilon_p(\br,t) =
p^2/2m + U(\br,t)$, where $U(\br,t)=U_{\rm ext}(\br) + 2 g [n_c(\br,t)
+ \tilde{n}(\br,t)]$. We expect this semi-classical description to
break down only for very low temperatures where the Bogoliubov
excitation spectrum is more appropriate~\cite{Giorgini97}. Our entire
discussion is within what is called the Popov approximation in that we
have ignored all effects associated with the anomalous pair
correlations $\tilde m(\br,t) = \langle \tilde \psi(\br,t) \tilde
\psi(\br,t) \rangle$.

The coupled equations~(\ref{eq1})-(\ref{eq3}) have been used to derive
the generalized two-fluid hydrodynamic equations in the
collision-dominated region described by a local-equilibrium Bose
distribution~\cite{Zaremba99,Nikuni2000}. They also have been recently
used to give a detailed analysis of condensate growth by quenching the
thermal cloud distribution~\cite{Bijlsma2000}. In these papers,
(1)-(3) are solved with both the condensate and non-condensate being
treated dynamically and allowed to be out of equilibrium. The key
limitation of the present paper is that we only consider the dynamics
of the condensate, with the thermal cloud being in static
equilibrium. This assumption allows a simple theoretical development
and should be adequate for out-of-phase modes. The key role of the
condensate and non-condensate being out of diffusive equilibrium was
first stressed in a series of papers by Gardiner and
coworkers~\cite{Gardiner2000,Davis2000}. These were based on a kinetic
master equation formalism quite different from what we use, and no
application was made to the damping of condensate collective modes.

It is important to understand what is meant by the
{\emph{collisionless regime}} and to clarify how this terminology
relates to the present work. Above $T_{\rm{BEC}}$ (where $C_{12}=0$),
equation \eq{3} reduces to the Boltzmann equation describing a normal
gas~\cite{Kavoulakis98,Guery-Odelin1999}. In this case, the
collisionless region is well defined and corresponds to having
$\omega_i \tau_{\rm{cl}} \gg 1$, where $\omega_i$ is the collective
mode frequency of the gas, on the order of the trap frequency, and
$\tau_{\rm{cl}}$ can be approximated by the mean time between
collisions described by the classical Boltzmann collision integral
$C_{22}$. In static equilibrium, this collision rate for a uniform gas
is given by \beq {1\over\tau_{\rm{cl}}} = \sqrt{2} n \sigma \bar{v},
\label{eqtcl} \eeq where $n$ is the density of atoms, $\sigma=8\pi
a^2$ is the quantum collision cross section, and $\bar{v}$ is the
average speed of an atom in the gas. Both above and below
$T_{\rm{BEC}}$, the analogous collision time corresponding to
collision processes described by $C_{22}$ in \eq{3} will give an
estimate of the lifetime of a {\emph{single-particle}} excitation in
the thermal cloud. This is distinct from the physics given by
$C_{12}$, which describes the collisions of condensate atoms with
atoms from the thermal cloud. In particular, one finds that if the
thermal cloud is described by the equilibrium Bose distribution
$(f=f^0)$, then $C_{22}[f^0,\Phi]=0$ but $C_{12}[f^0,\Phi] \neq
0$. Thus $C_{12}$ will give rise to damping of condensate
oscillations even when the thermal cloud is treated statically. Of
course, in the collisionless region, there is another source of
damping arising from the dynamical mean-field coupling between the
condensate and thermal cloud that is also included in \eq{1} and
\eq{3}; this is Landau
damping~\cite{Szepfalushy74,Shi97,Pitaevskii97,Fedichev98a,Giorgini98,Guilleumas2000,Morgan2000,Reidl2000},
which will be discussed below.

\subsection{Static Popov approximation}
In the present work, we use these equations to calculate the damped
normal modes of the condensate given by the solutions of \eq{1}
assuming that the non-condensate {\mbox{atoms}} {\mbox{always}} remain
in static thermal equilibrium. For our model, this means we take \beq
f(\bp,\br,t) \simeq f^0(\bp,\br) = {1 \over {e^{\beta[p^2 /2 m +
U_0(\br) - \tilde{\mu}_0]}-1}} \,, \label{eq4} \eeq where
$\tilde{\mu}_0$ is the equilibrium chemical potential of the
non-condensate and $U_0(\br)=U_{\rm{ext}}(\br) + 2g[n_{c0}(\br) +
\tilde n_0(\br)]$. The detailed analysis given by ZNG shows
that the Bose-Einstein distribution in \eq{4} is a stationary solution
to \eq{3} when the condensate and non-condensate are in diffusive
equilibrium, which requires $\tilde{\mu}_0 = \mu_{c0}$, where
$\mu_{c0}$ is the equilibrium chemical potential of the condensate as
described by \eq{1}.

Using our finite $T$ ``static Popov'' approximation, equation \eq{1}
can be simplified to \beq i\hbar {\partial \Phi \over \partial t} = \Big [
- {\hbar^2 \over 2 m} \nabla^2 + U_{\rm{ext}} + g n_c + 2g \tilde{n}_0
- i \hbar R_0 \Big ] \Phi ,\label{eq1b} \eeq which describes the
condensate motion within the static thermal cloud. Here $\tilde{n}_0$
is the equilibrium density of the non-condensate and the damping term
$R_0$ is calculated using \beq \Gamma_{12}^0(\br,t) \equiv \int {d\bp
\over (2 \pi \hbar)^3} C_{12}[f^0(\bp,\br),\Phi(\br,t)].\label{eq2b}
\eeq Notice that $\Gamma_{12}^0(\br,t)$ depends on time now only
through $\Phi(\br,t)$.  Using the explicit general expression for
$C_{12}$ given in \mbox{Eq.~(23b)} of ZGN$^\prime$, one finds (see
also Ref.~\cite{Gardiner2000}) \beq \Gamma_{12}^0(\br,t) = {n_c(\br,t)
\over \tau_{12}(\br,t)} \big [e^{-\beta (\tilde{\mu}_0 -
\varepsilon_c(\br,t))} - 1 \big ],
\label{eqx1}
\eeq where we have defined the $C_{12}$ collision time \bea {1 \over
{\tau_{12}(\br,t)}} &\equiv& {2 g^2 \over (2\pi)^5\hbar^7} \!\int
\!\!d{\bf p}_1 \!\int \!\!d{\bf p}_2 \!\int \!\!d{\bf p}_3
\delta(\bp_c+{\bf p}_1-{\bf p}_2-{\bf p}_3) \nonumber \\ &\times&
\delta(\varepsilon_c+\tilde \varepsilon_{p_1} -\tilde
\varepsilon_{p_2}-\tilde \varepsilon_{p_3}) (1+f_1^0) f_2^0 f_3^0 \,.
\label{eqx2}
\eea Here the condensate atom local
energy is $\varepsilon_c(\br,t) = \mu_c(\br,t) + {1\over 2}m
v_c^2(\br,t)$ with the non-equilibrium condensate chemical potential
\beq \mu_c(\br,t) = -{\hbar^2{\bbox\nabla}^2\sqrt{n_c}\over
2m\sqrt{n_c}} + U_{\rm ext} +gn_c + 2g\tilde{n}_0\,.
\label{eqx3}
\eeq The condensate atom momentum is $\bp_c = m \bv_c$, and $f_i^0 =
f^0(\br,\bp_i)$. We have introduced the usual condensate velocity
defined in terms of the phase $\theta$ of the condensate $\Phi(\br,t)
= \sqrt{n_c(\br,t)} e^{i \theta(\br,t)}$ as ${\bf v}_c =
\hbar\bbox{\nabla}\theta(\br,t)/m$. A closed set of equations for
$\Phi(\br,t)$ is given by \eq{1b} and its complex conjugate combined
with \eq{x1} - \eq{x3}.

We note that in terms of $n_c$ and $\bv_c$, \eq{1b} is completely 
equivalent to the coupled equations~\cite{Zaremba99}
\bea
{\partial n_c \over \partial t} + \bbox{\nabla}\cdot(n_c{\bf v}_c)&=& 
-\Gamma_{12}^0[f^0,\Phi]  \nonumber \\
m\left({\partial\over\partial t}+{\bf v}_c\cdot 
\bbox{\nabla}\right) \bv_c&=&-\bbox{\nabla}\mu_c\, .
\label{eqnew1}
\eea It is easy to see from \eq{x1} that when the condensate is in
equilibrium with the thermal cloud according to {\mbox{$\mu_c
\rightarrow \mu_{c0} = \tilde{\mu}_0$}}, $\Gamma_{12}^0(\br,t)$ then
vanishes. It is clear that the description of the system given by
equations \eq{4} and \eq{1b}, or equivalently \eq{new1}, is valid only
if the condensate is slightly perturbed from equilibrium and the
condensate motion is essentially uncoupled from that of the thermal
cloud, which we can then treat statically. In order to describe the
condensate oscillations about equilibrium, we use the quantum
hydrodynamic variables $n_c(\br,t)=n_{c0}(\br)+\delta n_c(\br,t)$ and
$\bv_c(\br,t) = \delta \bv_c(\br,t)$, where $n_{c0}(\br)$ is the
equilibrium density of the condensate with the associated equilibrium
chemical potential $\mu_{c0}$. Alternatively, one may work with the
fluctuations of $\Phi(\br,t)$ and derive coupled Bogoliubov
equations~\cite{Dalfovo99,Dodd98,Hutchinson97} generalized to include
the effect of the $R_0$ damping term. This generalization will be
discussed elsewhere~\cite{Williams2000}.

\subsection{Finite-$T$ Stringari wave equation}
From \eq{new1}, we can obtain linearized equations of motion for the
condensate fluctuations $\delta n_c$ and $\delta \bv_c$. We use the
fact that, to lowest order in the fluctuations from static
equilibrium, \eq{x1} reduces to \beq \delta \Gamma_{12}^0 = {\beta
n_{c0}(\br) \over {\tau_{12}^0(\br)}} \delta \mu_c(\br,t) \, , \eeq
where the ``equilibrium'' $C_{12}$ collision rate is defined by \bea
{1\over {\tau^0_{12}(\br)}} &\equiv& {2 g^2 \over (2\pi)^5\hbar^7}
\int d{\bf p}_1 \int d{\bf p}_2 \int d{\bf p}_3 \delta({\bf p}_1-{\bf
p}_2-{\bf p}_3) \nonumber \\ &\times&\delta
\big({p_1^2-p_2^2-p_3^2\over{2m}}-g n_{c0}\big ) (1+f_1^0) f_2^0 f_3^0
\,.\label{eq7} \eea In the present paper, we restrict ourselves to the
Thomas-Fermi limit, valid for large $N_c$, \bea {\partial \delta n_c
\over \partial t} + \bbox{\nabla}\cdot(n_{c0}\delta \bv_c)&=& -{1
\over \tau'}\delta n_c
\label{eq5} \\ m {\partial \delta \bv_c \over\partial t}
&=&-g\bbox{\nabla} \delta n_c \,. \label{eq6} \eea The collision time
$\tau'(\br)$ describes collisions between the condensate and
non-condensate atoms when the condensate is perturbed away from
equilibrium, \beq {1\over\tau'(\br)} = {g n_{c0}(\br) \over
{k_{\rm{B}} T}} {1 \over {\tau^0_{12}(\br)}} \, . \label{eqx4} \eeq In
the TF approximation the equilibrium distribution reduces to $f_i^0 =
[e^{\beta(p_i^2 /2 m + g n_{c0})}-1]^{-1}$. The new term on the
right-hand side of \eq{5} causes damping of the condensate
fluctuations due to the lack of collisional detailed balance between
the condensate and the static thermal cloud. We note that this
collision time is only a function of $\br$ through its dependence on
the static condensate density $n_{c0}(\br)$. Plots of $1/\tau'(\br)$
will be discussed below.

We can easily combine \eq{5} and \eq{6} to obtain what we shall refer
to as the finite $T$ Stringari wave equation \beq {\partial^2 \delta
n_c \over \partial t^2} - { g \over m} \bbox{\nabla}\cdot (n_{c0}
\bbox{\nabla} \delta n_c) = -{1 \over \tau'} {\partial \delta n_c
\over \partial t} \,.
\label{eq8}
\eeq Equation~(\ref{eq8}) is the main result of this paper. If we
neglect the right-hand side, we obtain the undamped finite $T$
Stringari normal modes $\delta n_c(\br,t) = \delta n_i(\br) e^{-i
\omega_i t}$ given by the solution of~\cite{Stringari96} \beq - { g
\over m} \bbox{\nabla}\cdot \big[n_{c0}(\br) \bbox{\nabla} \delta
n_i(\br)\big] = \omega_i^2 \delta n_i(\br).
\label{eq9}
\eeq As has been noted by several authors in recent
{\mbox{papers}~\cite{Giorgini98,Guilleumas2000,Dodd98}, $n_{c0}(\br)$
at finite $T$ can be well approximated by the TF condensate profile at
$T=0$ but with the number of atoms in the condensate $N_c(T)$ now
being a function of temperature, since the static mean field of the
non-condensate plays such a minor role. With this approximation for
$n_{c0}(\br)$, the solutions of the finite $T$ Stringari
equation~(\ref{eq9}) will be identical to those at $T=0$, since the
$T=0$ Stringari frequencies do not depend on the magnitude of
$N_c$. Of course, as shown by calculations solving the coupled
Bogoliubov equations~\cite{Hutchinson97,Dodd98}, the TF
approximation breaks down when $N_c \lesssim 10^4$. Thus the
condensate collective mode frequencies will always become temperature
dependent close to $T_{\rm{BEC}}$, where the TF approximation is no
longer valid. We generalize our present discussion to deal with this
region in Ref.~\cite{Williams2000}.

We can use the undamped Stringari modes as a basis set to solve \eq{8}
and find the damping of these modes. Writing $\delta n_c(\br) = \sum_i
c_i \delta n_i(\br)$, and using the orthogonality condition $\int d
\br \delta n_i(\br) \delta n_j(\br) = \delta_{ij}$, one obtains the
following algebraic equations for the coefficients $c_i$ \beq \omega^2
c_i = \omega_i^2 c_i - i \omega \sum_j \gamma_{ij} c_j \,,
\label{eq11}
\eeq
where 
\beq
\gamma_{ij} = \int d\br \delta n_i(\br) \delta n_j(\br) /
\tau'(\br)
 \, .
\label{eq12}
\eeq Assuming the damping is small (we are in the collisionless
region), \eq{11} is easily solved using perturbation theory by setting
$\gamma_{ij}=0$ for $i\neq j$. This gives the damped Stringari
frequency (to lowest order) $\Omega_i = \omega_i - i \Gamma_i$, with
\beq \Gamma_i \equiv {\gamma_{ii} \over 2} = {1 \over 2} \int d \br {
\delta n_i(\br)^2 \over \tau'(\br)}.
\label{eq13}
\eeq This result for $\Gamma_i$ is reasonable, namely it involves an
average over $1/\tau'(\br)$ weighted with respect to the undamped
density fluctuations of the Stringari wave equation \eq{9}. We find
that coupling to other modes ($\gamma_{ij}\neq 0$) is extremely
small.

\section{Results}

\subsection{Homogeneous gas}\label{homogen}
Before treating the trapped gas, it is useful to first apply our
theory to a homogeneous gas, which was considered previously in
Ref.~\cite{Nikuni99} in connection with the collision-dominated
hydrodynamic region. For a homogeneous gas, $\tau'$ is independent of
position and then \eq{13} reduces to $\Gamma_i = 1/2\tau'$. Although
our model in the present paper applies only to the collisionless
region, it is useful to compare the inter-component collision time in
both the collisionless and hydrodynamic regimes. In ZNG, it
was shown that the inter-component collision time $\tau_{\mu}$ in the
hydrodynamic region is given by $\tau_{\mu}=\sigma\tau'$, where the
temperature-dependent factor $\sigma$ (not to be confused with the
collision cross section) depends on various thermodynamic
functions. In Fig. 1 we compare $1/\tau'$ and $1/\tau_{\mu}$ as
functions of $T$. We see that $\sigma$ dramatically alters the
inter-component relaxation rate $1/\tau_{\mu}$ appropriate to the
hydrodynamic regime, as compared to $1/\tau'$ involved in the
collisionless regime. For completeness, in Fig. 1 we also plot the
often-used classical collision time given by \eq{tcl} as well as
$\tau_{12}^0$ defined in \eq{7}.

In a uniform Bose gas at finite temperatures, the Landau damping
($\omega = c q - i\Gamma_L$) of condensate modes has been evaluated in
several recent
papers~\cite{Shi97,Pitaevskii97,Fedichev98a,Giorgini98,Guilleumas2000,Morgan2000,Reidl2000}. Working
within the full second-order Beliaev approximation, one finds \beq
\Gamma_L = \Big ( {3 \pi \over 8} \Big ) {a k_{\rm{B}} T q \over
\hbar} \, .
\label{eqL1}
\eeq This is clearly quite different from our inter-component damping
$\Gamma = 1/2\tau'$, as plotted in Fig. 1. Landau damping originates
from the interaction of a condensate collective mode with the
excitations of the thermal cloud but is not associated with $C_{12}$
collisions which give rise to $\tau'$. 
\begin{figure}
  \centerline{\epsfig{file=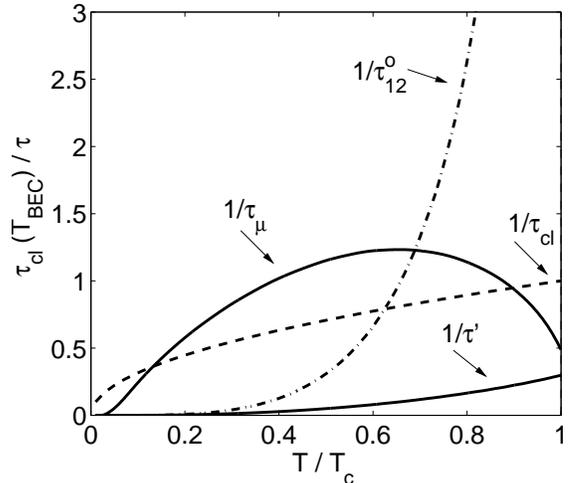,height=2.6in}}
\caption{Collision rates in a homogeneous Bose-condensed gas,
normalized to the collision rate of a classical gas at the BEC
transition temperature $\tau_{\rm{cl}}^{-1}(T_{\rm{BEC}})$. We have
taken $gn = 0.1 k_{\rm{B}} T_{\rm{BEC}}$, where $n$ is the total
density. See also Refs.~\protect\cite{Nikuni99,Griffin97b}.}
\end{figure}

In the context of our generalized GP equation in \eq{1}, Landau
damping comes from the fluctuations in the thermal cloud induced by
the condensate mean field, \beq \delta \tilde{n} = \tilde{\chi}_0(2 g
\delta n_c) \, . \label{eqL2}\eeq In the finite temperature region of
interest, $\tilde{\chi}_0$ can be approximated as the density response
function of a non-interacting gas of atoms with a spectrum
$\tilde{\varepsilon}_p$ and chemical potential $\mu_{c0}$. For a
uniform gas, one sees that using \eq{L2} in \eq{1}, with $R=0$, gives
condensate modes satisfying $\omega^2 = c^2 q^2 (1 + 4 g
{\tilde\chi}_0)$ and thus $\omega = c q - i \Gamma_L$, where \beq
\Gamma_L = 2 g c q \imag {\tilde \chi}_0(q,\omega=c q) \,
.\label{eqL3} \eeq Evaluating $\imag {\tilde\chi}_0$ in the limit of
small $q$ \cite{Szepfalushy74}, one finds $\Gamma_L = {4\over 3} a
k_{\rm{B}} T q/\hbar$. Apart from the slightly larger numerical
coefficient, this agrees with the exact result given in equation
\protect\eq{L1}~\cite{comment}.

\subsection{Trapped gas}
We now turn to explicit calculations of the inter-component damping
rate using our model for a trapped gas. In order to calculate
$\Gamma_i$ for a given mode, the equilibrium chemical potential
$\mu_{c0}=\tilde\mu_0$ must be calculated self-consistently for a
given total number $N$ of atoms at a given temperature $T$. In the TF
approximation, the procedure is straight forward (see, for example,
Ref.~\cite{Minguzzi97a}). In the following, we consider a harmonic
trap with axial symmetry $U_{\rm{ext}}(\br) = {1\over 2}m
\omega_{\rho}^2(\rho^2 + \lambda^2 z^2)$, where $\lambda
=\omega_z/\omega_{\rho}$ is the anisotropy parameter. In the TF
approximation, the condensate density takes the explicit form
$n_{c0}=[\mu_{c0}-m\omega_{\rho}^2(\rho^2 + \lambda^2 z^2)/2]/g$
within the TF radius, and the condensate chemical potential is
$\mu_{c0}={1 \over 2} \hbar \omega_{\rho} [15 \lambda N_c
a/\rho_0]^{2/5}$, where $\rho_0=\sqrt{\hbar/m \omega_{\rho}}$. The
form of the Stringari normal modes $\delta n_i(\br)$ is given
explicitly in the literature~\cite{Stringari96,Dalfovo99}.  We mainly
consider the breathing mode ($n=1,l=0$) for which
$\omega_{10}=\sqrt{5} \omega_\rho$, for $\lambda=1$.

We choose experimentally accessible parameters in the following
calculations for the collisionless region. However, we do not compare
our results to the two available experiments on damping of normal
modes at finite $T$, since the TF approximation is not valid for most
of the JILA data~\cite{Jin97}, and the MIT
experiment~\cite{Stamper-Kurn98} is approaching the
collision-dominated hydrodynamic limit where the dynamics of the
condensate and non-condensate become more strongly coupled. For
${}^{87}$Rb the scattering length is $a\simeq5.7$
nm~\cite{Hall98b}. We first consider a spherically symmetric trap
$\lambda=1$, with trap frequency $\nu_r = 10$ Hz, and we take
$N=2\times 10^6$. In the collisionless limit, we require $\omega_i
\tau_{\rm{cl}} \gg 1$, taking $\tau_{\rm{cl}}$ as defined in \eq{tcl}.
For a trapped gas, we obtain an upper limit on $1/\tau_{\rm{cl}}$ by
taking the density in the center of the trap $n(r=0)$, which gives
$1/\tau_{\rm{cl}} =8a^2N\omega_{\rho}^3 m /(\pi k_{\rm{B}}T)$. For the
parameters we use, $\omega_{10} \tau_{\rm{cl}} \approx 19$ (compared
to $\omega_{02} \tau_{\rm{cl}} \approx 20$ for the JILA
data~\cite{Jin97}, and $\omega_{02} \tau_{\rm{cl}} \approx 2$ for the
MIT data~\cite{Stamper-Kurn98}).

\begin{figure}
  \centerline{\epsfig{file=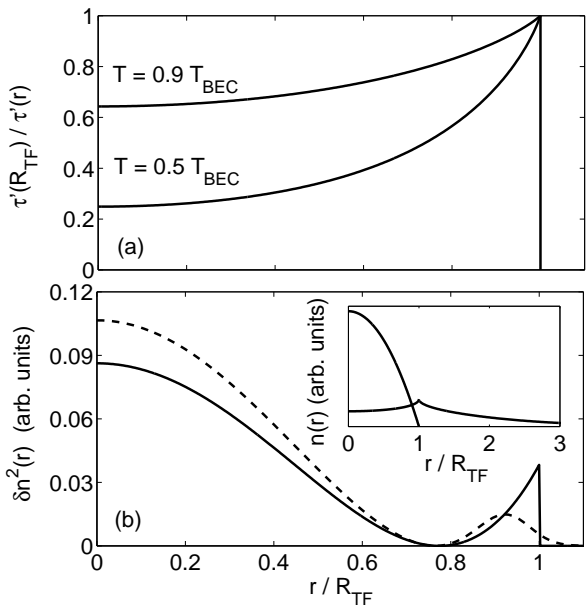,height=3.2in}}
\caption{Positional dependence of various quantities. In (a) we plot
$1/\tau'(\br)$ normalized by its value at the TF radius $R_{\rm{TF}}$.
In (b) we show the density fluctuation $\delta n_{10}$ of the
Stringari breathing mode (solid) for a spherically symmetric trap. We
also show the exact $T=0$ Bogoliubov mode (dashed) for $N_c(T=0.9
T_{\rm{BEC}})= 2.3 \times 10^5$. Both solutions are normalized to
unity, $\int \delta n_{10}^2(\br) d\br = 1$. The densities of the
condensate and thermal cloud are plotted in the inset for
$T=0.9T_{\rm{BEC}}$.}
\end{figure}
In Fig. 2a we plot $1/\tau'(\br)$ vs. position for $T=0.9T_{\rm{BEC}}$
and $T=0.5T_{\rm{BEC}}$. We see that the collision rate increases
steadily up to the condensate boundary, but as $T$ increases,
$1/\tau'(\br)$ becomes relatively constant. The behavior of
$1/\tau'(\br)$ just seems to be mimicking the behavior of the
non-condensate density $\tilde n(\br)$, which we plot in the inset of
Fig. 2b along with the condensate density. The condensate mean-field
pushes the non-condensate density out of the center of the trap, a
well known result~\cite{Dalfovo99}. We also show the breathing mode
density fluctuation in Fig. 2b. The sharp cusp of $\tilde n(\br)$ and
the sudden drop of $\delta n_i(\br)$ and $1/\tau'(\br)$ at the
condensate boundary are all unphysical artifacts of the TF
approximation. Inclusion of the kinetic energy pressure in a more
accurate calculation would have the effect of smoothing out this
behavior at the boundary. To illustrate the effect of the kinetic
energy pressure, we also show in Fig. 2b the breathing mode obtained
by solving the $T=0$ coupled Bogoliubov equations. We estimate that an
improved treatment which includes the kinetic energy pressure will
modify our estimate of $\Gamma_{10}$ by about
$10-20\%$~\cite{Williams2000}.

In Fig. 3a we plot the damping rate $\Gamma_{10}$ for the breathing
mode ($n=1$, $l=0$) shown in Fig. 2b as a function of temperature up
to $T=0.95 T_{\rm{BEC}}$, where $N_c\approx7\times 10^4$. At higher
temperatures, the Thomas-Fermi approximation will start to break down
and the mode frequencies become temperature
dependent~\cite{Hutchinson97,Dodd98}.
\begin{figure}
  \centerline{\epsfig{file=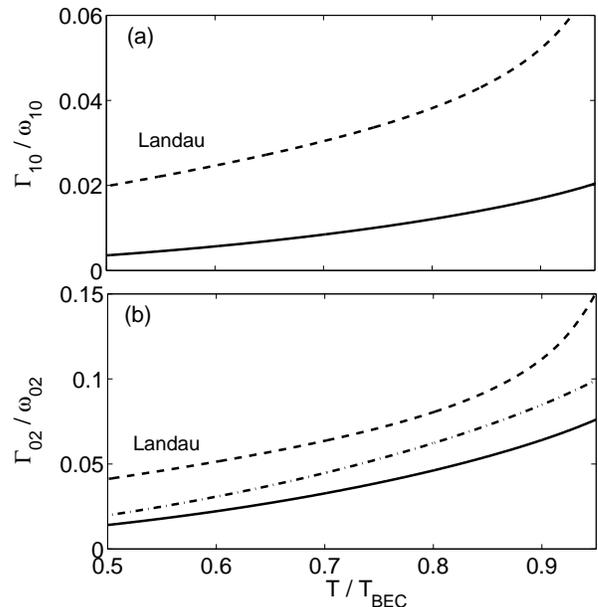,height=3.2in}}
\caption{Normal mode damping rates vs. temperature. In these plots,
the damping rates are normalized by their corresponding mode
frequencies and we only plot up to $T=0.95T_{\rm{BEC}}$, above which
the Thomas-Fermi approximation will start to break down. In (a) we
show the damping rate for the breathing mode ($n=1$, $l=0$) of a
spherically symmetric trap, where the solid line corresponds to
inter-component collisional damping given in \eq{13}. In (b) we show
damping rates for the quadrupole modes ($n=0$, $l=2$) in a cylindrical
trap.  The solid line is for the $m=0$ mode and the dot-dashed
line is for $m=2$.}
\end{figure}

Landau damping of condensate modes in trapped gases has also been
discussed in some detail in recent
papers~\cite{Fedichev98a,Guilleumas2000}. These papers give results
that are in qualitative agreement with the expression for a uniform
gas in \eq{L1}, with $q = \omega_{\rho}/c$ and evaluating the
Bogoliubov sound velocity $c$ for the density at the center of the
trap~\cite{Pitaevskii97}. This simple estimate for Landau damping is
plotted in Fig. 3 for comparison. We note that it is larger but
comparable to the inter-component collisional damping that we
consider. Clearly, a fully satisfactory theory of finite $T$ damping
of normal modes must include {\emph{both}} Landau damping as well as
the damping we consider in this paper due to the condensate being out
of diffusive equilibrium with the non-condensate.

Our theory is easily applied to anisotropic traps. In Fig. 3b we show
the damping of $m=0,2$ quadrupole modes for an axially symmetric trap
with $\lambda = \sqrt{8}$. Here we choose a slightly tighter trap
$\nu_r = 23$ Hz, and we take $N=1 \times 10^6$ (in this case,
$N_c\approx 3\times 10^4$ at $T=0.95 T_{\rm{BEC}}$).  For these
parameters, we find $\omega_{20} \tau_{\rm{cl}} \approx 6$. In Fig. 3b
we see that Landau damping is about twice as large as our
inter-component collisional damping.

\begin{figure}
  \centerline{\epsfig{file=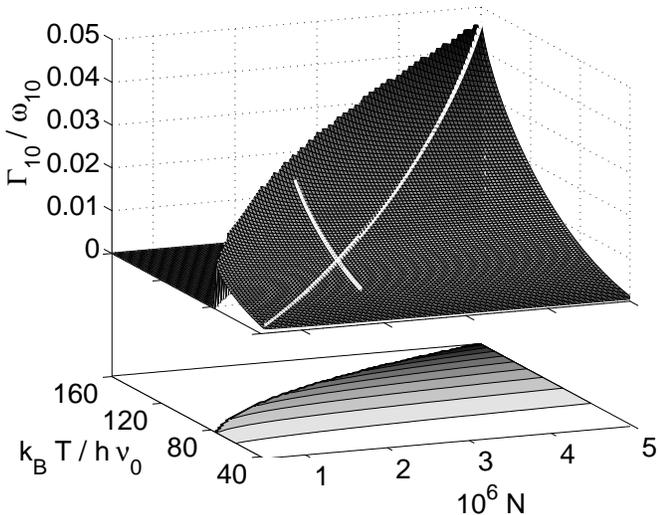,height=2.8in}}
\vspace{.2cm}
\caption{Damping rate vs. $T$ and total $N$ of the breathing
mode. Here we plot the surface of $\Gamma_{10}$ and its contours
projected onto the plane below. The solid white line at $N=2\times
10^6$ corresponds to the solid line plotted in Fig. 3a. The longer
line illustrates that in experiments, $N$ decreases as the
temperature is lowered due to evaporative cooling. The upper edge of
the surface corresponds to the critical temperature
$(k_{\rm{B}}T_{\rm{BEC}}/\hbar\omega_{\rho})=0.94 N^{1/3}$, above
which $\Gamma_{10}$ vanishes.}
\end{figure}
It is instructive to also consider the dependence of the mode damping
$\Gamma_i$ on the total population $N$. In Fig. 4, we show a shaded
surface plot of $\Gamma_{10}$ for the breathing mode in an isotropic
trap as a function of $T$ and $N$.  The white line at $N=2\times 10^6$
corresponds to the solid line plotted in Fig. 3a. As one might expect,
the inter-component damping rate increases with increasing total
population (since the density is increasing). It is also important to
realize that in current experiments, the data taken is for a broad
range of $N$ due to evaporative cooling
losses~\cite{Jin97,Stamper-Kurn98}. The idealized fixed-$N$ line can
never be achieved in practice and one is instead dealing with a curve
like the longer line on the surface in Fig. 4.

\section{Conclusion}
In summary, we have calculated a new damping mechanism of condensate
collective modes due to collisions with the thermal cloud, based on
the finite-$T$ equations derived in Ref.~\cite{Zaremba99}. The
essential mechanism involves the lack of diffusive equilibrium between
the condensate and the thermal cloud~\cite{Gardiner2000}, which also
plays a key role in the theory of condensate
growth~\cite{Bijlsma2000,Davis2000}. Here we have
carried-out the first explicit calculation of this damping mechanism
for a trapped gas in the collisionless regime (this inter-component
damping has recently also been evaluated in the collision-dominated
hydrodynamic regime~\cite{Zaremba2000}). In recent discussions of the
damping of condensate collective modes in the collisionless region,
the mechanism we consider is omitted. One instead focuses on the
dynamical mean-field coupling between the condensate and thermal
cloud, which gives rise to Landau and Beliaev
damping~\cite{Szepfalushy74,Shi97,Pitaevskii97,Fedichev98a,Giorgini98,Guilleumas2000,Morgan2000,Reidl2000}. While
we have not considered it in detail, we have indicated how we could
include Landau damping by considering the non-condensate fluctuations
in \eq{1} induced by the condensate mean
field~\cite{Giorgini98}. Comparing Landau damping to the additional
mechanism we have calculated, we find that the two are comparable in
size. Further experimental studies of the collective modes at finite
temperatures are needed to clarify the relative importance of these
different sources of damping.

In this paper, we have argued that a good first estimate of the
inter-component damping of condensate collective modes can be obtained
by coupling it to a static thermal cloud. A more systematic theory is
clearly desirable in which the collisionless dynamics of the thermal
cloud is allowed for. However, as noted in the introduction, we do not
believe that this will lead to significant corrections to the
inter-component damping of out-of-phase condensate modes in which the
motion of the thermal cloud is not significant. In future work, we
hope to discuss the damping due to $C_{12}$ collisions of collective
modes that mainly involve the motion of the thermal cloud, with the
condensate being treated statically.

This work grew out of discussions with E. Zaremba about the role of
$C_{12}$ collisions in the collisionless region. We also thank
T. Nikuni for useful comments. This research was supported by a grant
from NSERC.

\end{document}